\documentclass[isre,nonblindrev]{informs3}

\OneAndAHalfSpacedXII

\usepackage{natbib}
 \bibpunct[, ]{(}{)}{,}{a}{}{,}%
 \def\bibfont{\small}%
\TheoremsNumberedThrough     

\EquationsNumberedThrough    

\usepackage{graphicx}
\usepackage{subcaption}
\usepackage{tikz, threeparttable}
\usetikzlibrary{arrows.meta, positioning}
\usepackage{booktabs}
\usepackage{threeparttable}

\begin{document}




\TITLE{Does the AI Reviewer Alleviate or Aggravate the Maintenance Burden? Evidence from GitHub Copilot Reviewer}

\ARTICLEAUTHORS{%
\AUTHOR{Feiyang (Amber) Xu, Poonacha K. Medappa, Murat M. Tunc}
\AFF{Tilburg University, the Netherlands, \\\EMAIL{f.xu\_1@tilburguniversity.edu, p.k.medappa@tilburguniversity.edu, \\m.m.tunc@tilburguniversity.edu}, \URL{}}

} 

\ABSTRACT{%
  This study examines how AI-based code reviewers reshape contribution behavior and maintenance dynamics in open-source software (OSS) communities. AI reviewers 
  provide scalable, fast, and standardized evaluations. They 
  reshape contributions as well as maintainers' workload. Leveraging the introduction of GitHub Copilot Reviewer on GitHub as a natural experiment in 2025, 
  we investigate how AI review affects maintainers' workload and review efficiency as well as contribution behavior. We find that repository maintainers initially face a sharp increase in PR submissions
  and a higher review workload after the adoption of the AI reviewer. Over time, 
  human contributions decline while the increase in bot-generated PRs is sustained. Interestingly, survival analysis reveals that AI reviewers increase review efficiency for human PR contributions, 
  whereas bot PRs take longer to review and merge. Our findings suggest that while AI reviewers increase review efficiency, they aggravate the maintenance burden 
  by encouraging more code contributions via humans and bots.
}%



\maketitle

%


\section{Introduction}

While generative AI (GenAI) tools have substantially lowered the cost of producing highly polished outputs in knowledge-intensive industries \citep{cui2025effectsAIhighskilled, Lee2024}, they exacerbate information overload for maintainers, project managers, and reviewers, who must screen an expanding volume of submissions \citep{li2025exploring}. Traditionally, knowledge-intensive industries such as software development and academic research have relied on expert reviewers to assess and maintain the quality of knowledge work \citep{bana2026fightingfire, baumann2026stop}. However, this approach is increasingly under pressure because the influx of AI-assisted contributions has increased the maintenance burden on reviewers and project maintainers \citep{xu2025aiassistedprogramming, Demirer2026writingcode_shippingcode}.

Expert reviewers assess not only technical correctness, security, and maintainability, but also originality, aesthetic coherence, and collaborative potential. As GenAI models improve, delegating part of the evaluation to AI becomes possible. At the 2026 academic conference of the Association for the Advancement of Artificial Intelligence, which received more than 23,000 paper submissions, AI-generated reviews were produced within a day \citep{biswas2026aiassistedreview}. These reviews demonstrate that AI can perform reviews and evaluation at scale.

Compared to human reviews, AI reviews offer key operational advantages: they scale easily, process large volumes of submissions at near-zero marginal cost, provide near-instant feedback, and apply more standardized criteria that yield consistent evaluations. However, AI reviews differ from human reviews in evaluation depth \citep{biswas2026aiassistedreview}. Human reviewers rely on tacit knowledge, contextual understanding, and long-term architectural perspectives, whereas AI reviewers depend on pattern recognition and may struggle with novel problems, cross-module dependencies, and implicit design goals. Work shows that human reviews involve interaction, persuasion, mentorship, and leadership \citep{raimi2025judgmentalbot}, while AI feedback is suggestion-oriented and lacks negotiation and iterative clarification \citep{spiliotopoulou2026genaiwellbeing}. As a result, contributors may adapt by restructuring their work and breaking tasks into smaller units to align their contributions with AI evaluation \citep{baumann2026stop}. These adaptations suggest that, in the short run, automating the review process with AI may stimulate knowledge creation in the form of smaller, more incremental contributions. Further, evidence from code review bots in open-source software reveals a decrease in communication between project maintainers and contributors \citep{wessel2022quality}. Over time, this erosion of dialogue may weaken knowledge contributions in the long run. Whether AI reviewers ultimately diminish knowledge production therefore remains an open empirical question.

\begin{figure}[htbp]
\centering
\caption{Pull Request Workflow in OSS with Copilot Reviewer}
\label{fig:PR_diagram}

\begin{threeparttable}
\centering
\includegraphics[width=0.80\textwidth]{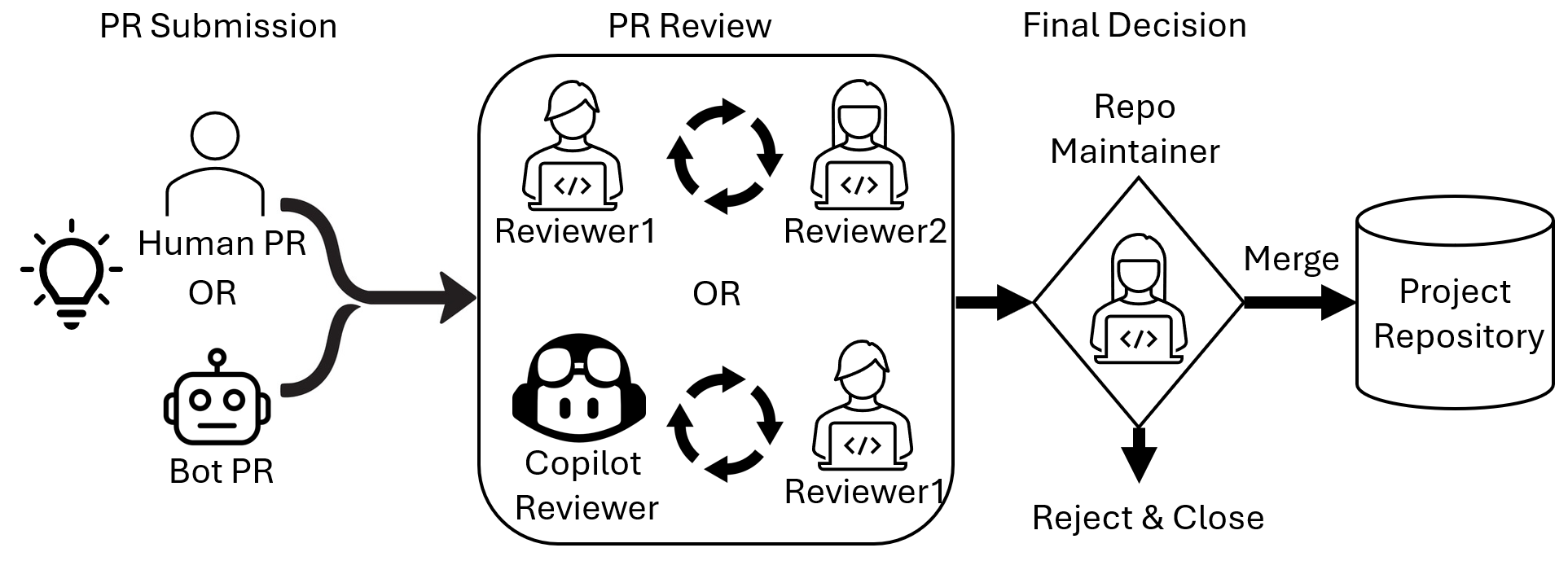}

\begin{tablenotes}[flushleft]
\footnotesize
\item Pull requests (PRs) can be submitted by human accounts or programmable bot accounts. The review process may take multiple rounds, involving Copilot Reviewer, human reviewers, or both. The final decision to merge or reject the PR is made by a human maintainer.
\end{tablenotes}

\end{threeparttable}
\end{figure}

This study examines how knowledge producers respond when part of the gatekeeping is taken over by AI. We leverage a natural experiment on GitHub, the world's largest open-source software (OSS) community, which recently introduced a GenAI-based pull request reviewer (Copilot Reviewer) into the code review process (see Figure \ref{fig:PR_diagram}). With GenAI integrated into the review process, the human maintainer of the repository can make a judgment on a pull request (PR) based on a short report generated by Copilot Reviewer. This workflow change allows us to observe how developers adjust their submission behavior when evaluative authority shifts from human peers to an algorithmic agent \citep{ostern2025bots_evaluate_humans}. The specific objectives of our research are:

\textit{How does the adoption of AI Reviewer impact knowledge production and maintenance in Open Source Communities?}

Our analysis reveals that the adoption of AI code reviewers initially improves knowledge production by human contributors. In the short run, human contributors respond positively, leading to a larger volume of PR submissions. Over time, however, human contributors decrease their PR submissions and increase the frequency of automated code generation via programmable bot accounts. Furthermore, we find that contributions made via humans become significantly smaller in size, while those from bot accounts are more comprehensive, with a higher number of additional lines of code. Our analyses also reveal that AI review increases the efficiency of the review process for PRs submitted by humans, while PRs submitted by bot accounts take longer to be successfully merged into the codebase. As the number of bot PRs increases, this may generate longer-term inefficiencies in knowledge maintenance. By situating AI as a gatekeeper for knowledge-intensive industries, this research contributes to the literature on AI-mediated review and evaluation processes. We provide empirical evidence on how shifting the locus of judgment — from human experts to AI reviewers — reshapes the behavior of knowledge workers in OSS communities and affects the maintenance burden.

\section{Literature Review}
\hfill
\subsection{GenAI in OSS Projects}

Recent advances in GenAI, exemplified by large language model (LLM) based tools such as ChatGPT and GitHub Copilot, are transforming both software production and work practices in OSS communities. Emerging research suggests that these tools affect not only how much code developers produce but also how they distribute their attention between coding, reviewing, and coordinating with others. For example, \cite{song2025impactgenerativeaicollaborative} find that GitHub Copilot adoption increases project-level code contributions but also increases the coordination time required for code integration. Similarly, \cite{hoffmann2025generative} show that access to GitHub Copilot reallocates developer effort toward core coding activities and away from project management and coordination. Together, these studies suggest that GenAI can expand the supply of code while shifting coordination demands to other stages of the development process.

This shift is particularly important because the ease of generating code does not eliminate the need for human validation. Although ``vibe coding'' can produce seemingly complete solutions with limited developer effort, AI-generated code may contain deficiencies that become visible only during subsequent review and integration. \cite{Schreiber_2025} show that even when AI-generated code passes functional tests, it can contain latent security vulnerabilities and maintainability problems, shifting effort from initial development toward downstream debugging and remediation. More broadly, the ease of producing AI-generated artifacts can create a new bottleneck: rather than generating contributions, OSS projects increasingly face the challenge of validating, prioritizing, and responding to them at machine-enabled scale \citep{Barrett2023}. This challenge has already become visible in practice. In January 2026, for example, the cURL project discontinued its bug bounty program after a surge of low-quality, AI-generated vulnerability reports consumed substantial maintainer attention while producing few actionable findings.\footnote{\url{https://www.techradar.com/pro/security/curl-will-stop-bug-bounties-program-due-to-avalanche-of-ai-slop}}

As GenAI increases the volume and speed of code production, OSS projects may therefore turn to automation on the review side as well: in effect, ``fighting fire with fire.'' Prior research provides initial evidence that automated gatekeeping can reshape OSS collaboration. \cite{wessel2022quality} study the adoption of rule-based code review bots on GitHub and find that bot adoption increases the number of merged PRs while reducing communication between maintainers and contributors.\footnote{Rule-based code review bots rely on deterministic workflow automation, such as dependency checks, linting, security scanning, and scheduled maintenance.} These bots, however, differ fundamentally from emerging GenAI reviewers. Traditional review bots apply predefined rules to specific tasks, whereas GenAI reviewers use LLMs to interpret code, generate contextual feedback, and perform review activities that more closely resemble evaluative judgments traditionally made by human maintainers. This transition from rule-based automation to AI-mediated evaluation raises a distinct question on how does AI reshape the behavior of contributors and the workload of human maintainers.

\subsection{GenAI for Feedback and Evaluation}

A growing literature examines the use of GenAI in settings traditionally governed by expert judgment, ranging from mental health screening to workplace evaluation and scientific peer review \citep{raimi2025judgmentalbot, kim2024mentalhealthAI, spiliotopoulou2026genaiwellbeing}. Across these settings, AI systems increasingly perform evaluative tasks by assessing individual inputs, providing feedback, and making recommendations that were previously the domain of human experts. The scalability and immediacy of AI evaluation create the potential to reduce expert workload and accelerate evaluation. Yet emerging evidence suggests that replacing or complementing human judgment with AI can also fundamentally change how individuals perceive and respond to the evaluation process.

One stream of research focuses on how individuals perceive AI evaluators and their authority. \cite{raimi2025judgmentalbot} show that conversational agents conducting mental health screening can be perceived as more judgmental than human professionals, even when the informational content of the interaction is held constant. Such perceptions reduce individuals' willingness to disclose information and engage with the evaluator, demonstrating that evaluation is not purely informational but also relational. Individuals' responses further depend on their perceived ability to interact effectively with AI. \cite{kim2024mentalhealthAI} show that the consequences of AI adoption depend critically on self-efficacy: individuals who are more confident in their ability to use AI experience more favorable outcomes, whereas those with lower self-efficacy are more susceptible to stress and reduced well-being. Beyond perception, individuals may also actively respond to AI's evaluative authority. \cite{ostern2025bots_evaluate_humans} show that individuals subject to supervisory bots contest and renegotiate algorithmic authority over time. Collectively, this literature suggests that introducing AI into evaluation changes not only who performs the evaluation but also how individuals perceive, engage with, and respond to the gatekeeper.

A second stream examines how AI-generated feedback changes individual effort and subsequent behavior. Although rapid and scalable feedback can improve access to evaluation, more feedback is not necessarily beneficial. Experimental evidence suggests that even positive AI-generated feedback can produce unintended consequences, including reduced effort, greater reliance on AI, and diminished intrinsic motivation \citep{spiliotopoulou2026genaiwellbeing, kinch2024negativeeffectsAIfeedback}. AI feedback can therefore alter the incentives surrounding knowledge production by changing how individuals interpret signals about the quality and progress of their work. Rather than simply improving an existing feedback process, AI evaluation may change the behavior that precedes and follows evaluation.

Taken together, this literature identifies two mechanisms through which GenAI can reshape expert evaluation. First, AI changes the perceived legitimacy and relational character of gatekeeping, thereby influencing how individuals engage with evaluators and respond to their authority \citep{raimi2025judgmentalbot, kim2024mentalhealthAI, ostern2025bots_evaluate_humans}. Second, AI-generated feedback changes behavioral incentives, potentially affecting effort, reliance on automated evaluation, and intrinsic motivation \citep{kinch2024negativeeffectsAIfeedback, spiliotopoulou2026genaiwellbeing}. These mechanisms are particularly consequential in knowledge-production communities, where contributors voluntarily decide whether, what, and how much to contribute. Thus, introducing GenAI into the evaluation process may affect not only the efficiency of reviewing existing contributions but also contributors' incentives and behaviors in producing new knowledge. These behavioral and perceptual consequences become especially salient when GenAI is embedded in institutionalized systems of expert evaluation.

\subsection{Infusing GenAI into Knowledge Maintenance}

As GenAI expands knowledge production, organizations face a corresponding challenge in evaluating and maintaining a growing volume of contributions. The managerial question is therefore shifting from whether to adopt AI toward how AI should be integrated into existing systems of expert evaluation. This tension is particularly visible in academic peer review, which, like OSS maintenance, relies on scarce expert attention. Journal Organization Science describes peer review as being under unprecedented pressure, with growing submissions creating unsustainable reviewer workloads \citep{gartenberg2026more}. GenAI offers a scalable response to this capacity constraint, but also raises questions about how automated evaluation should interact with human judgment.

One response is to move from opaque ``shadow AI,'' in which individual reviewers privately use general-purpose LLMs, toward institutionally governed AI. \cite{bana2026fightingfire} propose a human-in-the-loop system in which a centralized AI diagnostic precedes human review. Their model shows that governed AI can reallocate scarce reviewer attention and, under certain conditions, improve turnaround time and decision accuracy. This perspective positions GenAI not simply as an individual productivity tool, but as part of the institutional architecture of knowledge maintenance.

However, greater review capacity may create a trade-off between evaluation quality and utility. \cite{baumann2026stop} find that AI reviewers produce more homogeneous evaluations than human reviewers, reducing the diversity of judgment that peer review is designed to aggregate. They further show that AI review scores can be manipulated through superficial AI-assisted rewriting. Similarly, \cite{latona2024aireviewlottery} find that AI-assisted reviews tend to assign higher scores and can increase acceptance probabilities, particularly for submissions near decision thresholds. Thus, GenAI can affect not only the cost and speed of evaluation but also which knowledge is ultimately selected.

Together, these studies suggest that the value of GenAI depends critically on how it is positioned relative to human gatekeepers. Existing research largely examines AI as a substitute for human evaluation or as an assistant to individual reviewers. Much less is known about an increasingly relevant design: \textit{adding a centralized GenAI reviewer as an additional layer of evaluation while retaining human authority over final decisions}. Whether this arrangement reduces human maintenance burden or instead stimulates additional contributions and shifts more downstream work toward human experts remains an open question.

\section{Hypotheses Development}
\hfill
\subsection{Maintainer's Workload}

Automating part of the PR review process using AI gatekeepers may be perceived by developers as more impartial and less subject to social bias. The absence of interpersonal judgment and reputational concerns could reduce psychological pressure, encouraging experimentation and risk-taking. If developers view AI review as predictable and criteria-driven, they may rely on AI reviewer to calibrate their submissions, to improve measurable quality dimensions \citep{latona2024aireviewlottery}. In addition, once a project adopts a GenAI reviewer, contributors can expect instant and consistently available feedback, reducing uncertainty regarding the review process \citep{bana2026fightingfire}. This increased accessibility may lower the cost of participation and encourage a greater volume of PRs. Because final gatekeeping authority remains with human maintainers, we postulate the following in Hypothesis 1.

\textbf{Hypothesis 1} \textit{The adoption of GenAI Reviewer in the code review process increase maintenance load for maintainers.}

\subsection{Contribution Activities across Humans and Bots}
On GitHub, contributors make changes through PRs. Traditionally, PRs are authored by developers who understand the repository and submit well-scoped, context-aware contributions. Over time, repositories have also come to rely on rule-based bot accounts to open PRs on routine tasks. These bot accounts deterministically generate PRs in response to defined triggers, producing changes that require little contextual reasoning about the surrounding code.\footnote{LLM-driven agent bot accounts (e.g., GitHub's Copilot coding agent, see \url{https://docs.github.com/en/copilot/how-tos/use-copilot-agents/cloud-agent/create-a-pr}. ) are also becoming increasingly popular. However, majority of our data consists of PRs authored by rule-based bot accounts. According to \cite{Demirer2026writingcode_shippingcode}, agent bot can partially automating the PR submission process. Using this definition, we find that out of 119 bot account, 13 accounts can be classified as agent bot account, which count for 1.7\% of total PRs.} When an AI reviewer such as Copilot Reviewer is introduced, the response differs across these contribution types.

Prior research \citep{raimi2025judgmentalbot} in human-AI interaction suggests that individuals may perceive AI evaluators as more stringent or less forgiving than human professionals, even when objective criteria are identical. If developers anticipate harsher or less socially nuanced judgment from an AI reviewer \citep{spiliotopoulou2026genaiwellbeing, glickman2025human}, they may exhibit diminished intrinsic motivation and respond with greater caution — submitting more conservative, incrementally improved, or stylistically standardized code. 

Moreover, algorithmic reviewers may exhibit systematic affinities \citep{xu2025selfpreferencing}: in the context of hiring, LLMs consistently prefer resumes generated automatically over those written by humans. In the context of programming, GenAI reviewers may evaluate code generated by rule-based automated bots differently from human-written code, potentially leading to asymmetric scoring patterns. These effects may be particularly pronounced in OSS communities, where many contributors participate voluntarily (without organization control) and are often motivated by ideological commitments to openness, reputation, and community norms \citep{Daniel2018}. The change of code review standard, together with the tendency of LLMs to favor automatically generated content, will push contributors towards automating code generations via rule-based bot accounts. Hence, we formulate the hypothesis as follows.

\textbf{Hypothesis 2a} \textit{The adoption of GenAI Reviewer in the code review process reduce contribution by humans.}

\textbf{Hypothesis 2b} \textit{The adoption of GenAI Reviewer in the code review process increase contributions by rule-based bots. }

\subsection{Review Efficiency}

The introduction of AI-based reviewers fundamentally alters the speed and structure of the PR evaluation process on GitHub. Compared to human maintainers, LLM-based reviewers operate with near-zero marginal cost and can evaluate multiple PRs in parallel, substantially increasing review throughput. In addition, AI reviewers provide near-instant feedback, often within seconds or minutes, whereas human review is asynchronous and constrained by self-commitment. Together, these features reduce waiting times and alleviate bottlenecks in the review process, thereby improving overall workflow efficiency \citep{huang2026workflow}. Faster feedback cycles may also encourage contributors to iterate more quickly, potentially enhancing creativity and experimentation \citep{Lee2024}.

However, the efficiency gains from AI review are likely to be heterogeneous across types of contributions. For well-structured and coherent PRs—typically associated with human contributors—AI reviewers can accelerate evaluation and facilitate faster merging. In contrast, bot-generated PRs often involve larger, less focused changes that are harder to interpret and evaluate. For such submissions, the speed of AI feedback may not translate into effective communication or faster resolution, and may even exacerbate inefficiencies in the review process. Accordingly, we formulate the hypothesis as follows.

\textbf{Hypothesis 3a} \textit{PRs submitted by humans experience an increase in review efficiency after the adoption of GenAI reviewers.}

\textbf{Hypothesis 3b} \textit{PRs submitted by rule-based bots experience a decrease in review efficiency after the adoption of GenAI reviewers.}

\section{Methodology} 
\hfill
\subsection{Data}

In February 26, 2025, GitHub introduced Copilot Reviewer,\footnote{https://github.blog/changelog/2025-02-26-code-review-in-github-copilot-is-now-in-public-preview/} an AI-based tool embedded in the pull request workflow. This rollout generated plausibly exogenous variation in exposure to AI-assisted code review across repositories. Built on large language model (GenAI) capabilities, Copilot Reviewer evaluates submitted code by analyzing syntax, structure, and contextual intent, and generates natural-language feedback similar to that provided by human reviewers. Its core functions include identifying potential bugs, suggesting code improvements, flagging security vulnerabilities, and assessing adherence to coding standards.

\begin{figure}[htbp]
    \centering
        
    \begin{subfigure}{0.48\textwidth}
        \centering
        \frame{\includegraphics[width=\textwidth]{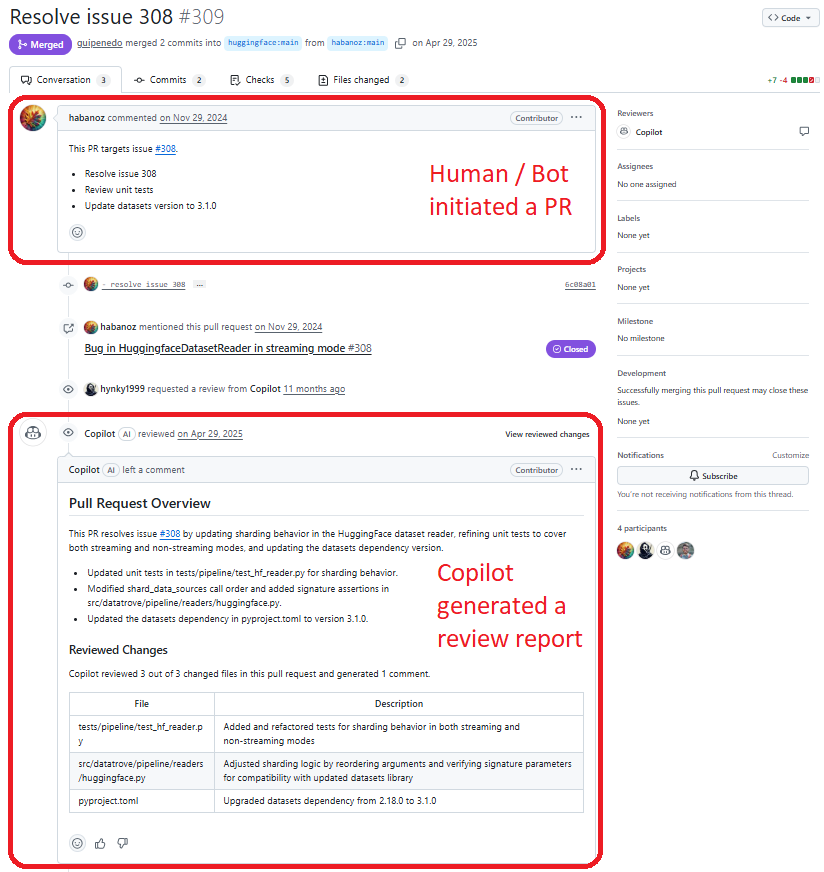}}
        \caption{Workflow of a successfully merged PR.}
    \end{subfigure}
    \begin{subfigure}{0.48\textwidth}
        \centering
        \frame{\includegraphics[width=\textwidth]{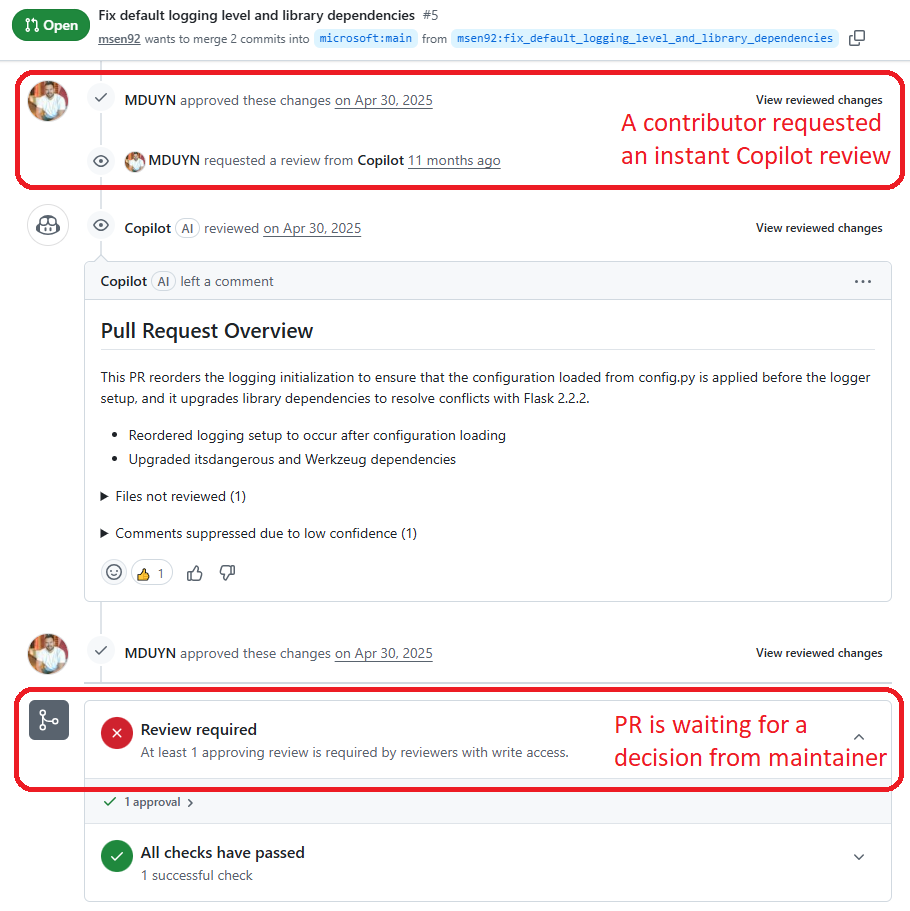}}
        \caption{Workflow of an open PR.}
    \end{subfigure}
    \caption{Screen short of how Copilot Reviewer is integrated into repository's teamwork}
    \label{fig:2x1graphs}
\end{figure}

Copilot Reviewer typically appears as an automated reviewer within the PR interface (see Figure 2a), where it can comment on specific lines of code, provide summary-level feedback, and recommend revisions. Unlike traditional rule-based tools (e.g., linters or continuous integration checks), which enforce explicit and codified constraints \citep{huang2026workflow}, Copilot Reviewer applies learned heuristics to approximate human judgment, enabling more flexible and context-sensitive evaluation.\footnote{https://docs.github.com/en/copilot/how-tos/use-copilot-agents/request-a-code-review/use-code-review}

The introduction of Copilot Reviewer alters the structure of the code review workflow. Traditionally, PR evaluation relies on human maintainers who provide iterative feedback and exercise discretion over acceptance decisions. With Copilot Reviewer, an initial layer of algorithmic evaluation is introduced prior to or alongside human review. This creates a hybrid workflow in which AI performs early-stage screening and feedback generation, while human reviewers focus on higher-level validation and decision-making (see Figure 2b).

To capture this workflow transformation, we utilize the Github's API service for data collection. Our empirical setting consists of repositories primarily written in Python that were initiated between July 2022 and February 2024. To ensure comparable maturity and ongoing development activity, we restrict the sample to repositories that remained active as of February 2026, had accumulated at least 100 stars, Python as the primary language, and has at least 10 PRs with at least 1 PR before and after the announcement of AI reviewer. Our dataset consists of 3,937 repositories.

\subsection{Identification Strategy}

Treatment status is defined at the maintainer level. A maintainer is classified as treated once “Copilot Reviewer” appears in a maintainer's PR review process. Based on this criterion, 689 maintainers are assigned to the treatment group.\footnote{\renewcommand{\baselinestretch}{0.5}\selectfont To address potential selection bias arising from repository owners' dual roles as contributors and maintainers, we conduct two alternative identification strategies: (1) excluding repository owners' maintenance activities in their own repositories and (2) excluding PRs submitted by repository owners to their own repositories. Both specifications focus on non-self-selected workload. Our results remain qualitatively similar under both approaches.} The control group comprises 1,902 maintainers with no observed Copilot Reviewer assisted PR review during the study period. Figure \ref{fig:rollout} graphically represents the number of treatment maintainers rollout during our observation period. 

\begin{figure}[htbp]
\centering
\includegraphics[width=0.95\textwidth]{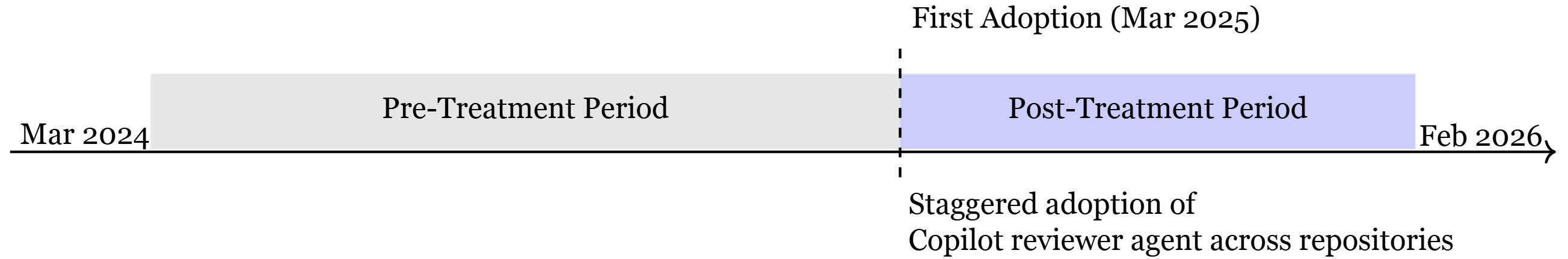}
    \caption{Timeline of Staggered Adoption of the Copilot Reviewer Agent}
    \label{fig:timeline_adoption}
\end{figure}

\begin{figure}[htbp]
\centering
\includegraphics[width=0.8\textwidth]{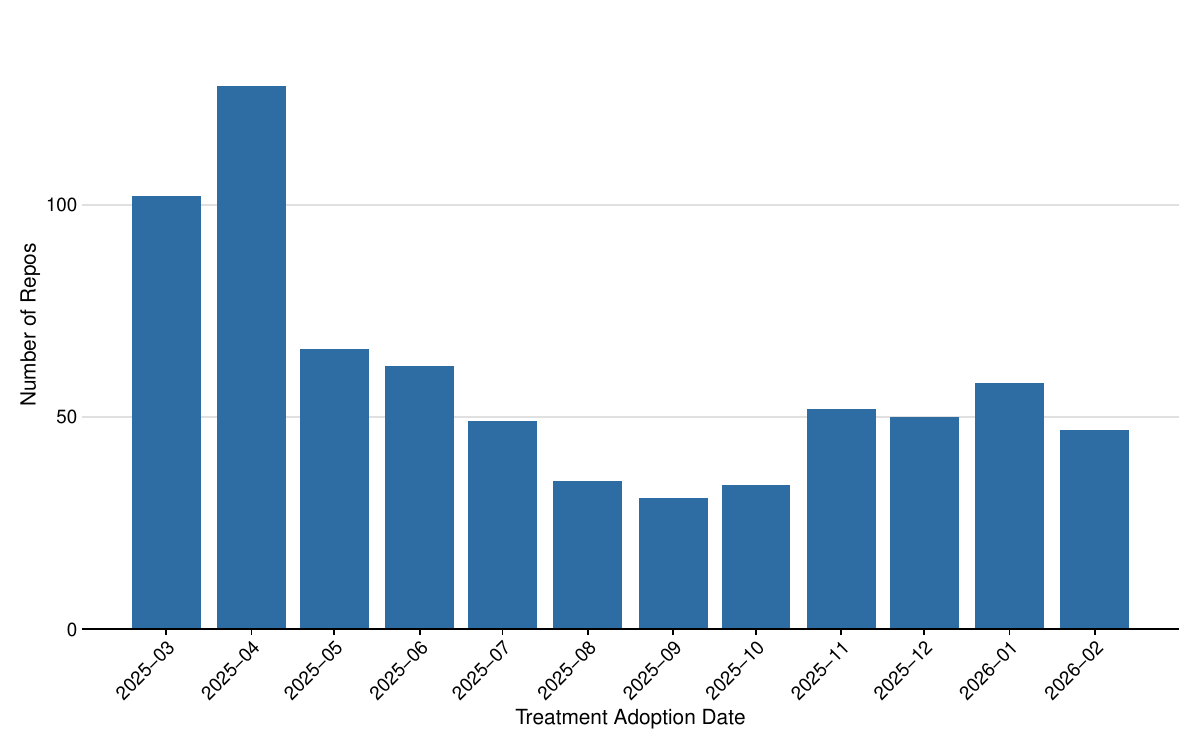}
    \caption{Treatment Rollout by Cohort for the Full Sample}
    \label{fig:rollout}
\end{figure}

Because Copilot Reviewer appears only when a PR requests review, we exclude the first Copilot-reviewed PR for each maintainers to mitigate adoption bias, using the timestamp of the Copilot Reviewer’s review action to identify initial exposure. We then examine PR-level performance. Our sample comprises approximately 974,654 PRs over the observation period.

\subsection{Variables}

To measure maintainer-level contributor behavior changes, we collected data on the number of PR and within PR teamwork activities. Using the GitHub API, we classify PR actors by type,\footnote{https://docs.github.com/en/enterprise-server@3.17/graphql/guides/using-global-node-ids} distinguishing between human users (“User”)\footnote{\renewcommand{\baselinestretch}{0.5}\selectfont We do not observe or classify the extent to which individual contributors use AI assistance. First, such use may occur as unobserved ``shadow AI'' across all account types. Second, our analysis takes the maintainer's perspective: regardless of whether AI assists in producing the code, a contribution submitted through a human account remains attributable to that account owner and carries the contributor's identity and reputation during the review process. We therefore classify contributions based on the observable account type (``User'' or ``Bot'') rather than the underlying degree of AI assistance.} and automated accounts (“Bot”). Our analysis explicitly differentiates between human- and bot-initiated activities.

\renewcommand{\arraystretch}{0.9} 
\begin{table}[htbp]
\centering
\caption{Descriptive Statistics}
\label{tab:descriptive_repo}

\footnotesize

\begin{threeparttable}
\begin{tabular}{p{3.6cm} p{7cm} r r r r r}
\toprule
Variable & Definition & Mean & SD & Min & Max & N \\
\midrule

AI Reviewer (Treatment)
& Dummy variable 1 if the maintainer has been assisted by the AI reviewer, 0 otherwise
& 0.089 & 0.285 & 0 & 1 & 62,184 \\

Reviewed Comments & Number of PR Review comments per PR checked by maintainer
& 0.790 & 0.870 & 0 & 5.485 & 62,184 \\ 

Maintained PRs & Number of PRs reviewed & 0.887 & 1.089 & 0 & 5.943 & 62,184 \\

Maintained Repos & Cumulative count of unique repositories reviewed
& 0.711 & 0.323 & 0 & 2.197 & 62,184 \\


Human PR Count 
& Number of PRs reviewed, authored by humans
& 0.845 & 1.068 & 0 & 5.943 & 62,184 \\

Bot PR Count 
& Number of PRs reviewed, authored by bots
& 0.084 & 1.068 & 0 & 4.419 & 62,184 \\

Additions [Human]
& Average lines of code added per human PR
& 4.478 & 0.375 & 0 & 14.028 & 62,184 \\

Changed Files [Human]
& Average number of files modified per human PR
& 1.771 & 1.009 & 0 & 9.828 & 62,184 \\

Commits [Human]
& Average number of commits per human PR
& 1.509 & 0.832 & 0 & 8.681 & 62,184 \\

Additions [Bot]
& Average lines of code added per bot PR
& 2.182 & 1.809 & 0 & 14.781 & 62,184 \\

Changed Files [Bot]
& Average number of files modified per bot PR
& 1.001 & 0.629 & 0 & 9.563 & 62,184 \\

Commits [Bot]
& Average number of commits per bot PR
& 0.805 & 0.408 & 0 & 6.768 & 62,184 \\

\bottomrule
\end{tabular}

\begin{tablenotes}
\footnotesize
\item \renewcommand{\baselinestretch}{0.5}\selectfont \textit{Notes:} The unit of observation is maintainer-month. All DVs are log-transformed. A value of one is added prior to transformation to address zeros.
\end{tablenotes}

\end{threeparttable}
\end{table}

At the PR level, we collected the data on PR merge durations. In the PR workflow on GitHub, each PR can end in one of three primary states: merged, closed, or open. A PR is \textit{merged} when the proposed code changes are accepted and integrated into the main codebase, indicating that reviewers deem the contribution sufficiently correct, useful, and aligned with project standards. A PR is \textit{closed} when it is rejected or abandoned without integration, which may reflect issues such as low code quality, misalignment with project goals, or insufficient revisions. A PR remains \textit{open} when it is still under review or awaiting further action, capturing ongoing evaluation and iteration. In line with our research questions, we only look at the successful code development: the merged PRs. 

These datasets enable us to examine the impact of the introduction of Copilot Reviewer on various aspects of project development and quality. The measures were aggregated to create a monthly panel dataset spanning from March 2024 to Feb 2026. Table \ref{tab:descriptive_repo} presents the main variables and provides descriptive statistics. Summary statistics are calculated based on repository-month observations.

\subsection{Model}

\subsubsection{Maintainer Level: DID}

To establish a causal relationship, we employ a staggered difference-in-difference (DID) approach at the maintainer level. This method allows us to isolate the effect of the Copilot Reviewer by comparing changes in behavior between treatment and control groups over time. By doing so, we can more confidently attribute any observed changes to the adoption of  Copilot Reviewer, rather than to other confounding factors. Following \cite{callaway2021difference}, the maintainer level staggered DiD model (repo \textit{i} month \textit{t}) is provided below:

\begin{equation}
\textit{Maintainer Level Effect}_{i,t} = \beta_{0} + \beta_{1}\textit{Copilot Reviewer}_{i,t} + \gamma _{i} + \delta_{t} +\epsilon_{i,t}
\end{equation}
where $\textit{Maintainer Level Effect}_{i,t}$ refers to the outcome measures for maintainer \(i\) in month \(t\). \(Copilot\ Reviewer_{i,t}\) is the independent variable and a binary indicator that turns to 1 when the Copilot Reviewer is adopted. The maintainer-level fixed effects are represented by \(\gamma_{i}\), and \(\delta_{t}\) represents the monthly fixed effects. \(\epsilon_{i,t}\) is the robust standard error clustered at the maintainer level to account for the potential heteroskedasticity of the errors.

\subsubsection{PR level: Survival Analysis}
At the PR level analysis, we focus on the time required for a PR to be merged, using survival analysis. In practice, not all PRs are merged within the observation window, and treating these incomplete observations as missing or excluding them would introduce selection bias. Survival models allow us to incorporate both completed (merged) and ongoing (unmerged) PRs while properly accounting for censoring. Moreover, the PR review process is dynamic, with the likelihood of merging evolving over time as feedback accumulates and revisions are made. The Cox proportional hazards framework is particularly suitable in this context, as it enables flexible estimation of how covariates affect the instantaneous probability of merge without imposing strong assumptions on the baseline hazard. This approach provides a more accurate and efficient characterization of PR workflow dynamics than static or cross-sectional models.

We estimate a Cox proportional hazards model to analyze how the hazard of merge evolves over the PR review process. Because PRs created in the final month of the observation window do not have sufficient time to realize a merge outcome, we exclude the last month of data to reduce right-censoring driven by the data boundary. In the model, a successful merge is treated as the event of interest, while PRs that are not merged during the observation window are treated as censored observations. To account for temporal variation in development activity and review conditions, we include year-month fixed effects. To control for time-invariant heterogeneity across repositories, we stratify the baseline hazard by repository. This specification allows us to compare merge dynamics across PRs while accounting for both repository-specific review environments and common time shocks.

\begin{equation}
h_{i,r,t}(\tau) = h_{0r}(\tau)\exp\left(
\beta_1 \,\textit{PRAuthorType}_{i,r,t}
+ \delta_t
\right),
\end{equation}

where
$h_{i,r,t}(\tau)$ is the hazard that PR $i$ in repository $r$ is merged at duration $\tau$ (unit in days),
$h_{0r}(\tau)$ is the repository-specific baseline hazard,
$\textit{PRAuthorType}_{i,r,t}$ is the PR author, either iniciated by human or bot,
$\delta_t$ denotes year-month fixed effects.
PRs not merged during the observation window are right-censored.

\section{Results}

Table \ref{tab:maintainer_workload} summarizes the estimations. We find that that maintainers review approximately 14\%\footnote{$(e^{0.131}-1)\times100 = 14\%$} more comments per PR and 10\% more PRs after treatment, while the number of repositories under their stewardship slightly decreases. Table \ref{tab:pr_authors} further shows that maintainers face a higher overall PR maintenance workload following treatment. Both human- and bot-initiated PR submissions increase significantly, by approximately 8\% and 5\%, respectively.

\begin{table}[htbp]
\centering
\caption{The Impact of AI Reviewer Adoption on Maintainer Workload}
\label{tab:maintainer_workload}

\begin{threeparttable}
\footnotesize

\begin{tabular}{lccc}
\toprule
Dependent Variable:
& (1) Reviewed Comments
& (2) Maintained PRs
& (3) Maintained Repositories \\
\midrule

AI Reviewer
& 0.131*** 
& 0.099***
& $-$0.006** \\

& (0.039)
& (0.036)
& (0.003) \\

\midrule

Maintainer FE
& Yes
& Yes
& Yes \\

Month FE
& Yes
& Yes
& Yes \\

Observations
& 62,184
& 62,184
& 62,184 \\

\bottomrule
\end{tabular}

\begin{tablenotes}[flushleft]
\footnotesize
\renewcommand{\baselinestretch}{1}\selectfont
\item \textit{Notes:} All DVs are log-transformed. Estimates are obtained using the Callaway and Sant'Anna (2021) difference-in-differences estimator with the control group defined as not-yet-treated maintainers. Robust standard errors clustered at the maintainer level are reported in parentheses. 
*** $p<0.01$, ** $p<0.05$, * $p<0.10$.
\end{tablenotes}

\end{threeparttable}
\end{table}
                            
\begin{table}[htbp]
\centering
\caption{The Impact of AI Reviewer Adoption on PR Authors}
\label{tab:pr_authors}

\begin{threeparttable}
\footnotesize

\begin{tabular}{lcc}
\toprule

& \multicolumn{2}{c}{Maintained Pull Requests} \\
\cmidrule(lr){2-3}

Dependent Variable:
& (1) Human-Authored PRs
& (2) Bot-Authored PRs \\
\midrule

AI Reviewer
& 0.079**
& 0.053*** \\

& (0.037)
& (0.015) \\

\midrule

Maintainer FE
& Yes
& Yes \\

Month FE
& Yes
& Yes \\

Observations
& 62,184
& 62,184 \\

\bottomrule
\end{tabular}

\begin{tablenotes}[flushleft]
\footnotesize
\renewcommand{\baselinestretch}{1}\selectfont
\item \textit{Notes:} All DVs are log-transformed. Estimates are obtained using the Callaway and Sant'Anna (2021) difference-in-differences estimator with the control group defined as not-yet-treated maintainers. Robust standard errors clustered at the maintainer level are reported in parentheses.
*** $p<0.01$, ** $p<0.05$, * $p<0.10$.
\end{tablenotes}

\end{threeparttable}
\end{table}

\begin{figure}[htbp]
    \centering
        
    \begin{subfigure}{0.495\textwidth}
        \centering
        \includegraphics[width=\textwidth]{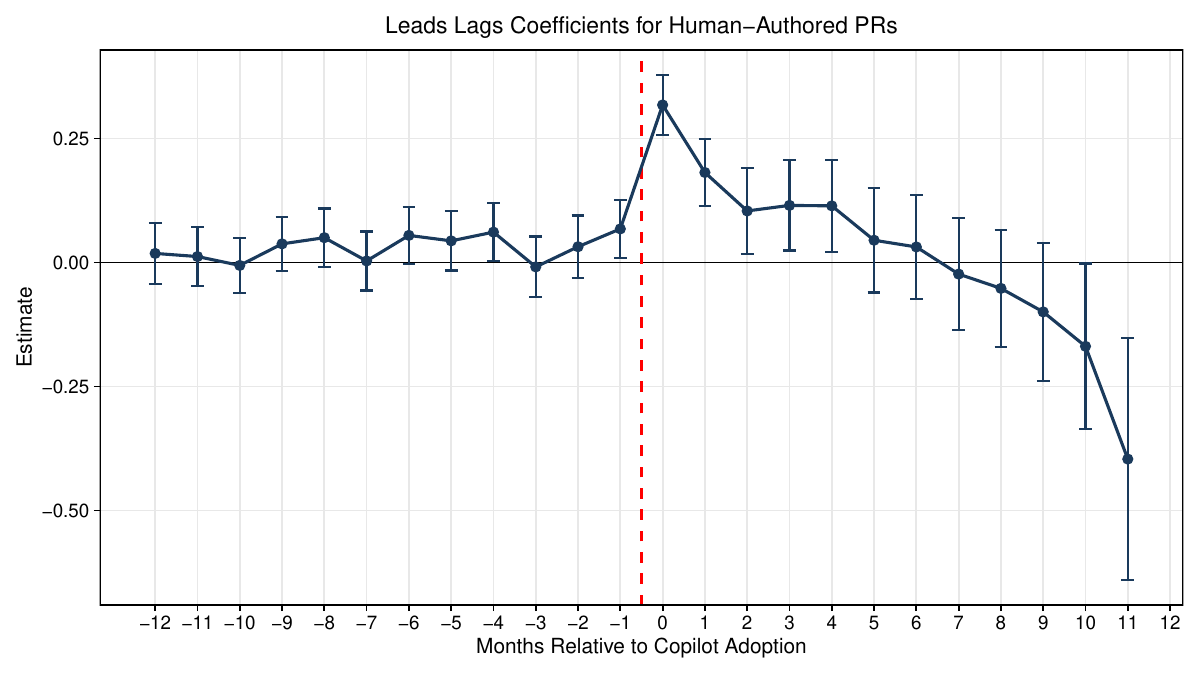}
        \caption{Human PR}
    \end{subfigure}
    \begin{subfigure}{0.495\textwidth}
        \centering
        \includegraphics[width=\textwidth]{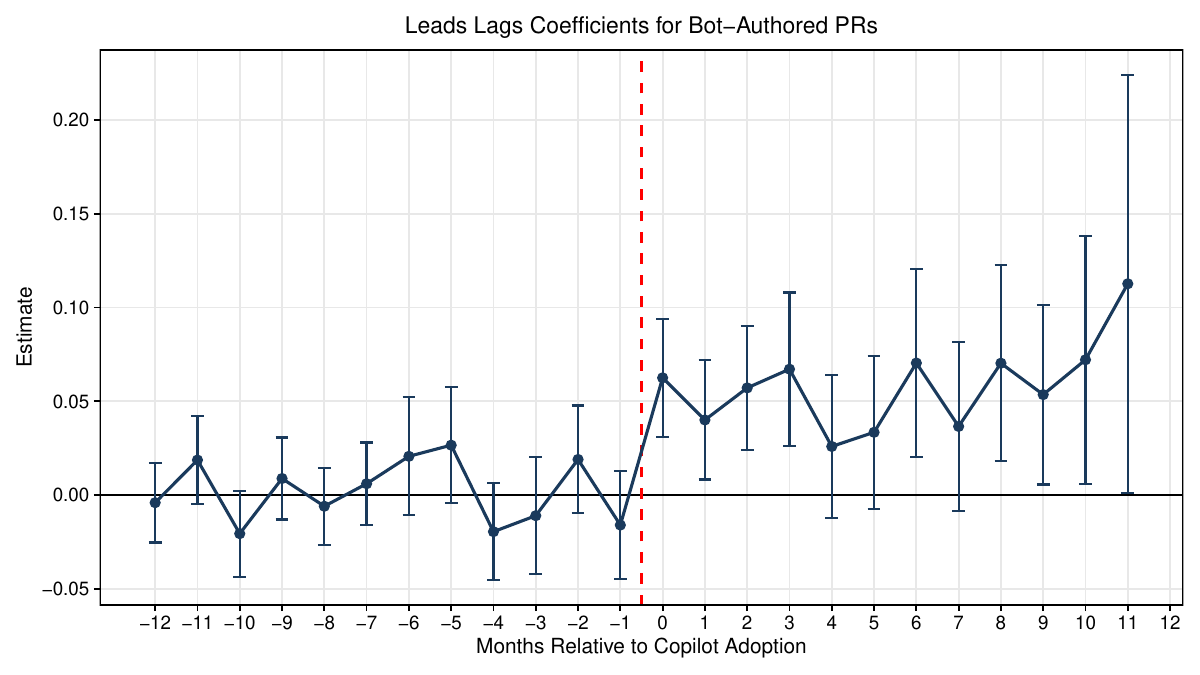}
        \caption{Bot PR}
    \end{subfigure}
    
    \label{fig:2x1graphs_pr}
    \caption{Plot of Lead and Lag Coefficients for Estimating the Impact of AI Reviewer on PR contributions}

\end{figure}

Examining the lead-lag dynamics for human and bot PRs (Figures 4a and 4b) reveals a more nuanced pattern. In the first month following the adoption of the AI reviewer, there is a substantial increase in human-initiated PRs. This surge suggests that repository contributors initially respond enthusiastically to the introduction of the AI reviewer and contribute more actively than in repositories without such a tool. However, this momentum dissipates quickly. After the initial spike, the number of human PRs declines to its pre-adoption level and eventually falls below that of the control group. In contrast, bot-initiated PRs exhibit a steady and sustained increase over time.

\begin{table}[htbp]
\centering
\caption{The Impact of AI Reviewer Adoption on the Characteristics of Development Activities}
\label{tab:pr_size_heterogeneity}

\begin{threeparttable}
\footnotesize

\begin{tabular}{lcccccc}
\toprule

& \multicolumn{3}{c}{Human-Authored PRs}
& \multicolumn{3}{c}{Bot-Authored PRs} \\

\cmidrule(lr){2-4}
\cmidrule(lr){5-7}

Dependent Variable:
& Additions & Files & Commits & Additions & Files & Commits \\

\midrule

AI Reviewer & $-$0.152** & $-$0.413*** & $-$0.173*** & 0.037** & 0.071 & 0.020 \\

& (0.061) & (0.145) & (0.066) & (0.015) & (0.044) & (0.021) \\

\midrule

Maintainer FE
& Yes & Yes & Yes
& Yes & Yes & Yes \\

Month FE
& Yes & Yes & Yes
& Yes & Yes & Yes \\

Observations
& 62,184
& 62,184
& 62,184
& 62,184
& 62,184
& 62,184 \\

\bottomrule
\end{tabular}

\begin{tablenotes}[flushleft]
\footnotesize
\renewcommand{\baselinestretch}{1}\selectfont
\item \textit{Notes:} Estimates are obtained using the Callaway and Sant'Anna (2021) difference-in-differences estimator with the control group defined as not-yet-treated maintainers. Robust standard errors clustered at the maintainer level are reported in parentheses.
*** $p<0.01$, ** $p<0.05$, * $p<0.10$.
\end{tablenotes}

\end{threeparttable}
\end{table}

Table 4 shows that human-initiated PRs decrease in size. All three measures—code additions $-$14.1\%, modified files $-$33.8\% and submitted commits $-$15.9\%, have significant negative coefficients. In contrast, bot-initiated PRs exhibit the opposite pattern. All three measures have positive coefficients, with average code additions increasing significantly by 3.8\%.

Taking the results in Tables 2, 3, and 4 and Figure 4 together, we find that for human-initiated PRs, Hypothesis 1 holds in the short run and Hypothesis 2a holds in the long run, as the graphical pattern indicates a spike in rising PR submissions by humans immediately after the introduction of Copilot Reviewer, followed by a decrease below the control group. For bot-initiated PRs, aligned with Hypothesis 2b, we can see both the volume and size of bot-initiated PRs increasing.

\begin{table}[htbp]
\centering
\caption{Cox Proportional Hazards Model: Time to PR Merge}
\label{tab:cox_merge}

\begin{threeparttable}
\footnotesize

\begin{tabular}{lccc}
\toprule
Variable & Coefficient & Hazard Ratio & SE \\
\midrule

Copilot Reviewed $\times$ Human PR
& 0.041***
& 1.042
& (0.011) \\

Copilot Reviewed $\times$ Bot PR
& $-$0.242**
& 0.785
& (0.078) \\

\midrule

Repository-stratified baseline hazards
& Yes & & \\

Month FE
& Yes & & \\

Observations
& 643,272 & & \\

Events (Merged PRs)
& 500,419 & & \\

Concordance
& 0.541 & & \\

\bottomrule
\end{tabular}

\begin{tablenotes}[flushleft]
\footnotesize
\renewcommand{\baselinestretch}{1}\selectfont
\item \textit{Notes:} The table reports estimate coefficients and hazard ratios from a Cox proportional hazards model estimating the time to PR merge. The baseline group is non-Copilot-Reviewed PRs. A hazard ratio greater than one indicates a higher likelihood of merging (shorter time to merge), while a ratio below one indicates a lower likelihood (longer time to merge). The model includes year-month fixed effects and repository-stratified baseline hazards. Standard errors are clustered at the repository level. $^{*}$\emph{p} $<$ .1; $^{**}$\emph{p} $<$ .05; $^{***}$\emph{p} $<$ .01
\end{tablenotes}

\end{threeparttable}
\end{table}

\begin{figure}[htbp]
\centering
\label{fig:survival_curves}

\begin{threeparttable}
\centering
\includegraphics[width=0.80\textwidth]{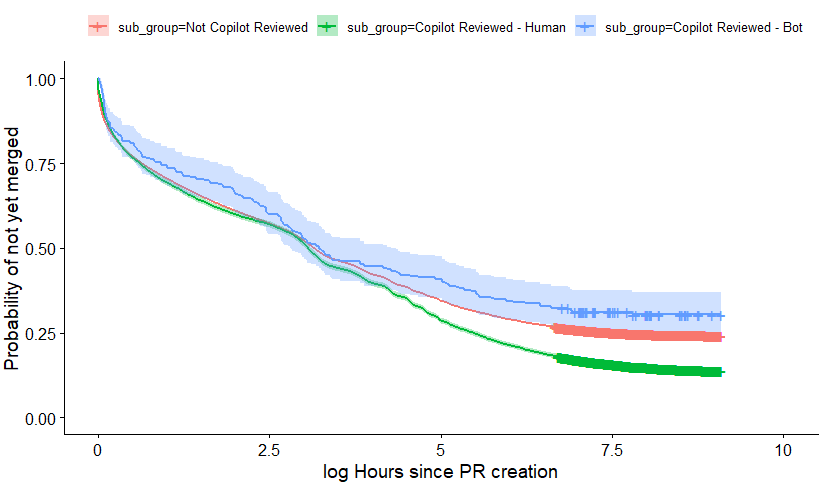}

\begin{tablenotes}[flushleft]
\footnotesize
\renewcommand{\baselinestretch}{1}\selectfont
\item \textit{Notes:} The survival function represents the probability that a PR remains unmerged over time. Although the statistical evidence Cox model uses days as the time unit, the x-axis in the figure is transformed to $\log_{10}(\text{hours}+1)$ to improve visualization, as many PRs are reviewed and merged within a few hours. For comparability, the sample is restricted to PRs initiated during the treated period; including pre-treatment PRs would results in the Not Copilot Reviewed group toward a longer durations, as some PRs remain open for the entire observation window. Shaded areas indicate 95\% confidence intervals.
\end{tablenotes}

\end{threeparttable}
\caption{Kaplan--Meier Survival Curves of Time to Pull Request Merge}

\end{figure}

Further evidence emerges when examining the heterogeneity of development activities using survival analysis among PRs not reviewed by Copilot Reviewer and human/bot PRs reviewed by Copilot Reviewer. In Table 5, we test Hypotheses 3a and 3b. PRs authored by humans and reviewed by Copilot Reviewer exhibit a higher hazard of merging (HR = 1.042), corresponding to an approximately 4.2\% increase in the instantaneous probability of merge relative to the baseline group (non-Copilot-reviewed PRs).

In contrast, PRs generated by bots and reviewed by Copilot Reviewer show a significantly lower hazard of merging (HR = 0.785), implying a 21.5\% reduction in the likelihood of merge. This indicates that bot-generated contributions, despite being subject to the same AI review process, are less likely to be successfully integrated and require longer evaluation periods, as shown in Figure 5. 

These results provide a proxy for contribution quality and review outcomes: a higher human PR merge rate relative to bot PRs suggests stronger alignment with quality expectations, whereas a lower bot PR merge rate indicates lower average quality or greater mismatch with project requirements. The share of not-successfully-merged PRs reflects workflow efficiency and review backlog, with prolonged openness potentially signaling increased complexity or reviewer overload.

\section{Mechanism}
\hfill
\subsection{PR Size Affected Review Efficiency}
To examine whether the speed advantage associated with Copilot review depends on the amount of code being evaluated, we estimate a Cox proportional hazards model that interacts Copilot review with PR size. Larger PRs contain more code for reviewers to understand and may therefore require greater contextual judgment, coordination, and verification. If Copilot review is most effective for smaller and less complex review tasks, its association with faster merging should weaken as the number of code additions increases. We measure PR size using the mean-centered value of $\log(1+\text{additions})$. The model uses the full PR sample, includes month fixed effects, allows the baseline merge hazard to vary across repositories, and clusters standard errors by repository. This interaction is a mechanism-relevant moderation test: it evaluates whether PR size conditions the relationship between Copilot review and merging speed, rather than establishing formal statistical mediation.

\renewcommand{\arraystretch}{0.9} 
\begin{table}[htbp]
\centering
\caption{Cox Proportional Hazards Model: Time to PR Merge}
\label{tab:cox_merge}

\begin{threeparttable}
\footnotesize

\begin{tabular}{lccc}
\toprule
Variable & Coefficient & Hazard Ratio & SE \\
\midrule
Copilot reviewed
    & $0.146^{***}$ & 1.157 &  (0.012) \\

Additions (mean-centered)
    & $-0.100^{***}$ & 0.904 & (0.001) \\

Copilot reviewed $\times$ Log additions
    & $-0.030^{**}$ & 0.971 & (0.005) \\

\midrule

Project FE
& Yes & & \\

Month FE
& Yes & & \\

Observations
& 643,272 & & \\

Events (Merged PRs)
& 500,419 & & \\


\bottomrule
\end{tabular}

\begin{tablenotes}[flushleft]
\footnotesize
\item \textit{Notes:} The dependent variable is the time from PR creation to merge. Unmerged PRs are right-censored. Log additions is defined as $\log(\text{additions}+1)$ and mean-centered. 
$^{***}p<0.001$, $^{**}p<0.01$, and $^{*}p<0.05$.
\end{tablenotes}

\end{threeparttable}
\end{table}

The results indicate that Copilot review is associated with faster merging for an average-sized PR. Because code additions is mean-centered, the main Copilot coefficient is evaluated at the sample mean of log additions. Its hazard ratio of 1.157 ($p<0.001$) indicates that, at this reference point, Copilot-reviewed PRs have a 15.7\% higher instantaneous merge rate than otherwise comparable non-Copilot-reviewed PRs. Larger PRs are generally merged more slowly: among non-Copilot-reviewed PRs, a one-unit increase in log additions is associated with a 9.6\% lower merge hazard (HR $=0.904$, $p<0.001$). More importantly, the interaction between Copilot review and log additions is negative and statistically significant (HR $=0.971$, $p=0.005$). Thus, each one-unit increase in log additions reduces the relative merge-hazard advantage associated with Copilot review by approximately 2.9\%. Equivalently, doubling additions reduces Copilot's relative hazard advantage by approximately 2.0\%. These findings suggest that Copilot review facilitates the integration of average-sized PRs, but its speed advantage becomes weaker as the amount of submitted code increases, consistent with the H3b that larger changes continue to require substantial reviewer attention and contextual judgment.

\section{Robustness Checks}
\hfill
\subsection{On and Off Switching}

Because the Copilot Reviewer is a subscription base service, and require repository-level configuration, and organization permissions / settings. Not every PR in the tratement repo is reviewed by Copilot Reviewer. In this section, we further analysis dynamic exposure patterns: (1) limited usage, where only a small portion of PRs are review by Copilot Reviewer, while majority of the PRs are remain maintainor reviewed, and (2) on-and-off switching, where maintainers repeatedly alternate between treated and untreated states.

\begin{table}[htbp]
\centering
\caption{The Impact of AI Reviewer Adoption on PR Authors, on and off switching}
\label{tab:pr_authors}

\begin{threeparttable}
\footnotesize

\begin{tabular}{lcc}
\toprule

& \multicolumn{2}{c}{Maintained Pull Requests} \\
\cmidrule(lr){2-3}

Dependent Variable:
& (1) Human-Authored PRs
& (2) Bot-Authored PRs \\
\midrule

AI Reviewer
& 0.216***
& 0.233*** \\

& (0.234)
& (0.102) \\

\midrule

Maintainer FE
& Yes
& Yes \\

Month FE
& Yes
& Yes \\

Observations
& 31,092
& 31,092 \\

\bottomrule
\end{tabular}

\begin{tablenotes}[flushleft]
\footnotesize
\renewcommand{\baselinestretch}{1}\selectfont
\item \textit{Notes:} All DVs are log-transformed. Estimates are obtained using the de Chaisemartin and D’Haultfoeuille (2024) difference-in-differences estimator. Robust standard errors clustered at the maintainer level are reported in parentheses.
*** $p<0.01$, ** $p<0.05$, * $p<0.10$.
\end{tablenotes}

\end{threeparttable}
\end{table}

We estimate dynamic treatment effects following the heterogeneity-robust framework of \cite{dechaisemartin2024continuousnostayers}. As an alternative to the binary treatment specification (ever treated versus never treated), we measure treatment intensity using the monthly \emph{Copilot penetration rate}, defined as the share of a maintainer's reviewed PRs that are also reviewed by Copilot Reviewer in a given month. This continuous measure captures variation in the intensity of Copilot Reviewer exposure rather than treating all exposed maintainers as equally treated. The findings suggest that the results remain consistent under the on and off switching identification: there is an respond spike by human-initiated PRs in the first month but dissipates quickly, falling below that of the control group; bot-initiated PRs exhibit a steady and sustained increase over time.

\section{Discussion}
\hfill
\subsection{Theoretical Contributions}

Our study highlights the hidden costs of introducing AI reviewers in software development within OSS communities. While AI code reviewers reduce the cognitive burden on repository maintainers, they may also inadvertently discourage direct human participation in the contribution process. As contributors increasingly delegate code generation and submission to automated agents, the knowledge production process becomes less straightforward. This shift makes it more difficult for maintainers to assess and justify the substantive value of code contributions, ultimately leading to higher rates of PR failure. To the best of our knowledge, this study represents the first empirical investigation of GenAI code review activities. Although prior literature has examined GenAI as a reviewer for academic papers \citep{bana2026fightingfire, biswas2026aiassistedreview}, software programming is more commercial and has broader applications in production. 

First, we provide empirical evidence and enrich relevant theories in information systems. We study a pioneering form of expert judgment conducted by GenAI. Contributors initially welcome the AI reviewer by submitting more PRs in the first few months, but this excitement fades, and contributions eventually drop to levels below those of control group. Second, we contribute to the literature on GenAI-integrated workflows. We find evidence of efficiency improvements when PRs are submitted by humans rather than bots. The GenAI reviewer is resolving the workflow bottleneck problem of lacking repo maintainers. Lastly, we find evidence that adoption of AI for code review may lead to blurred ownership over codebase quality in OSS. When repository owners adopt AI for code review, it signals a shift to contributors, and we observe more PRs being delegated to automated agents. However, bot-initiated PRs are less efficient and exhibit lower merge rates.

\subsection{Managerial Implications}

Our findings have important implications for knowledge-intensive work in the era of GenAI. As the capabilities of GenAI continue to improve, delegating expert feedback to AI systems is becoming increasingly common. This shift has two important implications. First, it increases the scalability and speed of the review process by reducing the marginal cost of evaluation. Second, it changes contributor behavior by making evaluation criteria more standardized and predictable. Contributors may adapt by tailoring submissions to satisfy algorithmic feedback or by delegating more code generation to AI tools. As a result, Copilot Reviewer not only augments the review process but also reshapes upstream production decisions, embedding algorithmic governance into the core of software development workflows. 

When GenAI enhances the capability of knowledge production, the central challenge shifts from scarcity to overabundance of generated information. This creates significant strain on evaluation processes, particularly in collaborative environments such as OSS development. As one repository maintainer noted after a very long PR review back and forth between a human author and Copilot Reviewer that led to an expansion of the PR’s scope:\footnote{https://github.com/Nerogar/OneTrainer/pull/1360}

\textit{“What is your plan regarding this PR? If you want to finish it towards a mergeable PR, I'd advise breaking it down into smaller pieces and submitting multiple PRs. It currently does multiple unrelated things… Also, please consider what this PR should achieve and what lies beyond its scope. It seems to attempt to do everything at once.”}

When programmers increasingly delegate code production to automated agents, the boundaries of tasks (operationalized as PRs in software development) become less clearly defined. This blurring of scope complicates evaluation and coordination, highlighting the need for more structured task decomposition and governance mechanisms. Our analysis shows that smaller human-initiated PRs are more likely to be merged. This suggests that, despite the productivity gains enabled by GenAI assistants, collaboration efficiency remains constrained by human cognitive limits. Therefore, the ability to keep code contributions at a scale that is manageable for human reviewers may represent an important dimension of collaborative efficiency in AI-assisted software development.

\subsection{Limitations and Future Research Directions}

Our study is not without its limitations. First, We need to identify the specific utility of AI assistants in the code review process. Automation tools have long been integrated into software development workflows on GitHub, making it difficult to isolate the distinct effects of GenAI. For example, tools like GitHub Actions \citep{huang2026workflow} automate routine tasks by enforcing explicit rules and standardizing quality assurance. In contrast, GenAI reviewers such as Copilot and Gemini externalize coding heuristics, substitute feedback loops, and reshape reviewer–contributor interactions. Second, our study focuses primarily on open-source coding. As GenAI continues to develop, the PR framework may become less suitable for emerging paradigms such as vibe coding. The unit of work submission may need to evolve to better fit new demands. Third, although our analyses of proxy variables (additions, files, and commits) for PR size are informative and provide suggestive evidence on code quality, we acknowledge that they may not capture all dimensions of PR quality.

There are several avenues for future research to extend this study. One potential direction is to explore the impact of GenAI as an expert judge in other knowledge production processes, such as healthcare or wealth management. Another interesting direction is to examine whether the adoption of agentic AI, or even AI-to-AI interaction affects knowledge production dynamics. We invite future researchers to extend along these directions.

\bibliographystyle{informs2014}
\bibliography{references.bib}


%
%
%




\end{document}